# Towards a Modular Recommender System for Research Papers written in Albanian

Klesti Hoxha, Alda Kika, Eriglen Gani, Silvana Greca
Department of Computer Science
University of Tirana, Faculty of Natural Sciences
Tirana, Albania

***Abstract*—In the recent years there has been an increase in scientific papers publications in Albania and its neighboring countries that have large communities of Albanian speaking researchers. Many of these papers are written in Albanian. It is a very time consuming task to find papers related to the researchers' work, because there is no concrete system that facilitates this process. In this paper we present the design of a modular intelligent search system for articles written in Albanian. The main part of it is the recommender module that facilitates searching by providing relevant articles to the users (in comparison with a given one). We used a cosine similarity based heuristics that differentiates the importance of term frequencies based on their location in the article. We did not notice big differences on the recommendation results when using different combinations of the importance factors of the keywords, title, abstract and body. We got similar results when using only the title and abstract in comparison with the other combinations. Because we got fairly good results in this initial approach, we believe that similar recommender systems for documents written in Albanian can be built also in contexts not related to scientific publishing.**

## I. Introduction

Relevant information retrieval is very important for the scientific community, but also a very time consuming task. The academic search engines usually use keywords to find the relevant articles. This approach often produces unsatisfying results. An alternative approach suggests the usage of a recommender system to facilitate the retrieval of relevant information [1] to potential users. A recommender system assists the users in the process of finding relevant and personalized information fast.

Recommender systems are designed to help users navigate through complex and overload information by suggesting which items a user could have interest [1]. They are used in many domains, including music, movies, TV programs, videos on demand, books, news, images, web pages, research papers etc. Their role in our information society is becoming essential. The interest in this area is high because recommender systems help in the process of localization of personalized information from the overload gathered data.

In Albania there are many scientific journals and conference proceedings, which produce a lot of scientific papers in Albanian language. The scientific papers are published in hard copy journals, optical media (i.e. CD-ROM) and in the corresponding journal web pages. The Albanian researchers usually search the articles related to their work through traditional digital means using specific keywords or even hand browsing the individual websites of particular journals or conferences. This labor-intensive task in searching for articles is often useless and the retrieved articles may not be the needed ones. The search task in the hard copy journals is even more exhaustive. Recommender systems can provide a considerate help to Albanian researchers to acquire proper information from digital libraries.

In this paper we propose the design of a system that considers the case of a scientific journal which has begun as a hard copy journal and now is in the process of creating a web application in order to publish and access the published papers which can be in the English and Albanian language. The fact that most of the scientific articles are written in Albanian makes the process more difficult since there are no concrete systems which deal with information retrieval tasks regarding documents written in Albanian. The proposed recommender system is designed in a modular form, enabling easy replacement and modification of the components and experimenting with new algorithms in the future.

To detect the similarity between the interest of users and the available resources different approaches can be used. Most of them make use of the vector space model [16] that represents the items in question as vectors in a vector space. Each dimension of this vector space represents a feature of the items. When calculating the similarity of items, a possibility is to calculate the distance between these vectors for example by using the cosine similarity described in [4]. When recommending items to users, they are usually ranked in decreasing similarity order in comparison to other vectors in the items vector space. The different recommendation approaches differ the way this vector (or a set of vectors) is chosen [1]. It may be an item that is known beforehand that is in the users interests, or it might consist of a set of preferences of the user stored as a vector in the same vector space of the items. Another suggested approach is to recommend to a particular user items liked by users with similar profiles to him [6].

The focus of this paper is to present the design of a system with a highly modular architecture and make the first steps towards a recommender system that recommends documents (scientific articles in our case) written in Albanian. We used a periodic scientific journal published in Albania as a testing dataset. It contains scientific papers written mostly in Albanian about five different research areas: mathematics, physics, biology, chemistry, and computer science. We did not have to crawl the web for collecting these articles because we had access to the articles' collection. For building the recommender system we made use of the cosine similarity measure which is generally used in various recommender systems for digital libraries [2, 3, 23]. The used dataset was relatively small; it consisted of 226 articles in total. However the aim of this work was to identify the experimentation settings that produce the better results with the chosen similarity measure.

Dealing with papers written in Albanian makes necessary the usage of a word stemmer [16] designed for this language. We used the Albanian stemmer suggested in [17] that has been successfully tested in a document classification context. In our recommender system we used this algorithm in two different ways: stemming words with a single run and stemming with multiple runs over the initial word. This experiment was performed because some Albanian words can be further reduced after the first run of the stemming algorithm proposed in [17].

We gained fairly good results in our experiments. There were relevant items in the list of articles recommended to the users. In the performed experiments, we differentiated the importance of each part of the article in the used similarity heuristics. Our results showed that there were no big differences among the produced results. It was demonstrated that the experiments that used only the abstract and title of the articles for the recommendation process produced better or same results as most of the other performed experiments. This result was gained also by Nascimento et al. [23]. In terms of computation this speeds up the recommendation process, because the processed text (title and abstract) consists of a small part of the article.

The achieved results produced positive insight about future improvements of the system, or the creation of new information retrieval systems that deal with nonscientific related documents (i.e. news articles, laws) written in Albanian.

The first part of this paper briefly presents some of the research literature related to the existing approaches of designing the recommender system. The other parts introduce the proposed system architecture, the technologies that have been used, and the similarity heuristics that have been tested. The paper is concluded by presenting the conclusions and future work.

## II. RELATED WORK

The three basic approaches used in the design of recommendation systems are: content-based, collaborative filtering and hybrid [1]. The content-based recommender systems [5] default strategy consists of matching up the previously collected attributes of a user profile with those of the items in question, with the intent to arrive at a relevant result. This comparison is usually done in a vector space that stores the items and the user profiles as feature vectors [16]. Each dimension of this vector space represents a particular item feature.

Another possible content-based recommendation system strategy analyzes item descriptions to identify items that are of particular interest to the users. The filtering techniques used in this approach rely on item descriptions and generate recommendations from items that are similar to those that the target user has liked in the past, without directly relying on the preferences of the users (stored in their profiles) or other individuals [5]. This last approach does not require a large user base and collected data about them. When lacking the latest, the collaborative-filtering approach (see below) would be ineffective. The content-based approach makes possible a recommendation based solely on the description of the items themselves and not the interested users. This is the case for the initial stage of most digital libraries and similar information retrieval systems [23]. The similarity comparison in this approach is straightforward because it compares an item with other items, however in order for the recommendation to make sense, there is the need for an initial item (article) that is known to be of some interest for the user.

Collaborative filtering recommender systems [6] recommend items based on the past preferences of similar users. The recommendation is based on the assumption that items liked by users with similar profiles to a concrete user, are highly probable to be liked by the latest. This requires having a solid user profile database that stores the preferences and activity related data about the users. If the users of the digital library are not actively participating by making reviews or providing some feedback about the articles, or if they do not have full specified profiles (research area, interests), this database would lack of important data for the recommendation process. However if these data exists, there is a high probability that the recommendation process produces good results [6], [7].

Hybrid recommender systems [7] usually use a combination of content based and collaborative filtering recommendation for recommending items. This combined approach deals with the drawbacks of the above described ones, allowing for an initial content-based recommendation in cases of a cold start (lack of user profiles) [23]. The collaborative-filtering recommendation can improve the results by adding context-related information to the content-based approach.

Although recommender systems are very popular in commercial applications these days, recommender systems for the academic research have also gained interest. This is noticed by the emergence of a lot of research papers about this topic presented in many conferences and journals. Below we describe some of the applications of recommender systems in scientific paper recommendation situations.

Docear is an academic literature suite to search, organize, and create research articles [8]. Its recommender system [9] uses content based filtering methods to recommend articles. It allows the users to build "mind maps" that represent a user model (profile) which is matched with Docear's Digital Library. The authors claim to have achieved decent results based on the number of clicks gained through about 30 thousand tested recommendation results.

In [10] a personalized academic research paper recommendation system is presented. It recommends relevant articles to the research field of the users. It is supposed that the users (researchers) "like" their own articles. Based on this assumption papers similar to the ones previously written by the system users are recommended as relevant to them. This system uses a web crawler to retrieve research papers from two concrete digital libraries: IEEE Xplore and ACM Digital Library. It uses text similarity to determine the similarity between two research papers and collaborative filtering methods to recommend the items.

Nascimento et al. provide another example of a content-based recommender system for scientific articles [23]. They point out that most of the recommender system approaches suppose that a large collection of scientific papers is available beforehand. This is the case for some digital libraries like IEEE Xplore, but it does not hold for many other situations. Their proposed solution depends on publicly available scientific metadata, concretely the title and abstract of the articles. Their designed system collects these data by simulating searches on the websites of various publishers. Instead of using user defined keywords, they generate keywords from a particular article that is presented by the users (most probably an article in their particular research area).

The similarity of the articles is calculated by using the cosine similarity based on the vector space model [4]. The same similarity measure is used in our designed recommender system (see below). The results gained by Nascimento et al. were fairly positive, demonstrating that it is enough to consider only the title and abstract of the articles for recommendation purposes.

The hybrid approach of recommender systems has also been used for recommending research papers [11, 12]. As an example we have Techlend, in which different techniques of combining content-based and collaborative-filtering recommendation algorithms have been compared [11]. The experiments were sequential ones. The results of one algorithm were fed to the other. In general this approach produced good results. The test dataset was quite large, including about 100 thousand research papers indexed in CiteSeerX[1]. In 85% of the cases, the users found at least one related article to the presented recommendation list. Because some of the performed experiments did not perform well, the authors suggest that the sequential execution of the two involved recommendation algorithms (content-based and collaborative-filtering) is not the best alternative.

Another approach used by some academic paper recommender systems uses the paper's citations for recommending articles. In [12] it is presented another hybrid recommendation system. It aimed to be a powerful alternative to academic search engines by not solely relying on keyword analysis, but by additionally using citation analysis, explicit ratings, implicit ratings, author analysis, and source analysis. The popular academic search engine CiteSeerX also uses citations to find similar scientific papers [26]. Some other applications with citation recommendation are presented in [13], [14], and [15].

To the best of our knowledge, there have been no serious works regarding intelligent recommender systems that deal with documents written in Albanian. Maybe the main reason was the lack of enabler tools written specifically for the Albanian language that are used by many information retrieval systems, i.e. word stemmers and part of speech taggers [16]. However, the situation seems to have changed recently. A few Albanian language stemming algorithms have emerged, for example the ones described in [17] and [18]. The same holds for part of speech (POS) taggers, as examples we have the ones proposed in [19] and [20].

The stemming algorithm proposed by Sadiku and Biba in [17] has been tested in classifying documents written in Albanian about biology, history, literary and chemistry. They noticed an accuracy increase when using the stemming algorithm in comparison when it was not used. They also pointed out that the results were worsened when classifying documents of related fields. This fact is not strictly connected to the stemming algorithm itself, biology and chemistry articles have similar words in their content.

## III. Context Description

In this paper we provide the design of an intelligent search system about academic papers written in Albanian. Right now it is very difficult for Albanian researchers to find works related to their field of research, present in several journals or conference proceedings published in Albania and its neighboring countries that have large communities of Albanian speaking researchers.

We aim to provide a system that not only allows for keyword based searching of scientific papers, but also recommends related articles (to a concrete article). In this case we suppose that the user "liked" a certain article, after finding it by a normal search (i.e. via keywords) and is interested in finding other articles similar to this one. The first requirement of our system is the creation of an index of articles that stores metadata about each of them and enables searching. The created index also allows for further investigations about a collection of articles written in Albanian, like topic identification [6] and research trends detection.

Our testing dataset was obtained from a periodical scientific journal published by the Faculty of Natural Science, University of Tirana, Albania[2]. It contains scientific articles about five main research fields: mathematics, physics, biology, chemistry, and computer science. Most of these articles are written in Albanian. They are provided in PDF format and follow some standard formatting guidelines (font weight, font size, etc.).

We aimed on increasing the visibility of each individual article of the above mentioned journal, allowing researchers to easily find scientific articles related to their research work. Also, because scientific articles have all a very similar structure, our implemented system can be used for other articles written in Albanian. This addresses a crucial need of the Albanian research community and would boost the quality of research work done in Albania and its neighboring countries.

[1] http://citeseerx.ist.psu.edu/

[2] The Bulletin of Natural Sciences (Buletini i Shkencave Natyrore), http://buletini.fshn.edu.al/

## IV. PROPOSED SYSTEM ARCHITECTURE

We designed a highly modular system architecture that allows for louse coupling of the proposed modules (shown in Figure 1). Each individual module is designed to be as much independent as possible, so we can easily extend or change the behavior of the actual system in the future. It is to a high degree independent of the actual scientific article formatting (it is possible to import multiple types of article layouts after a correct specification of the formatting rules).

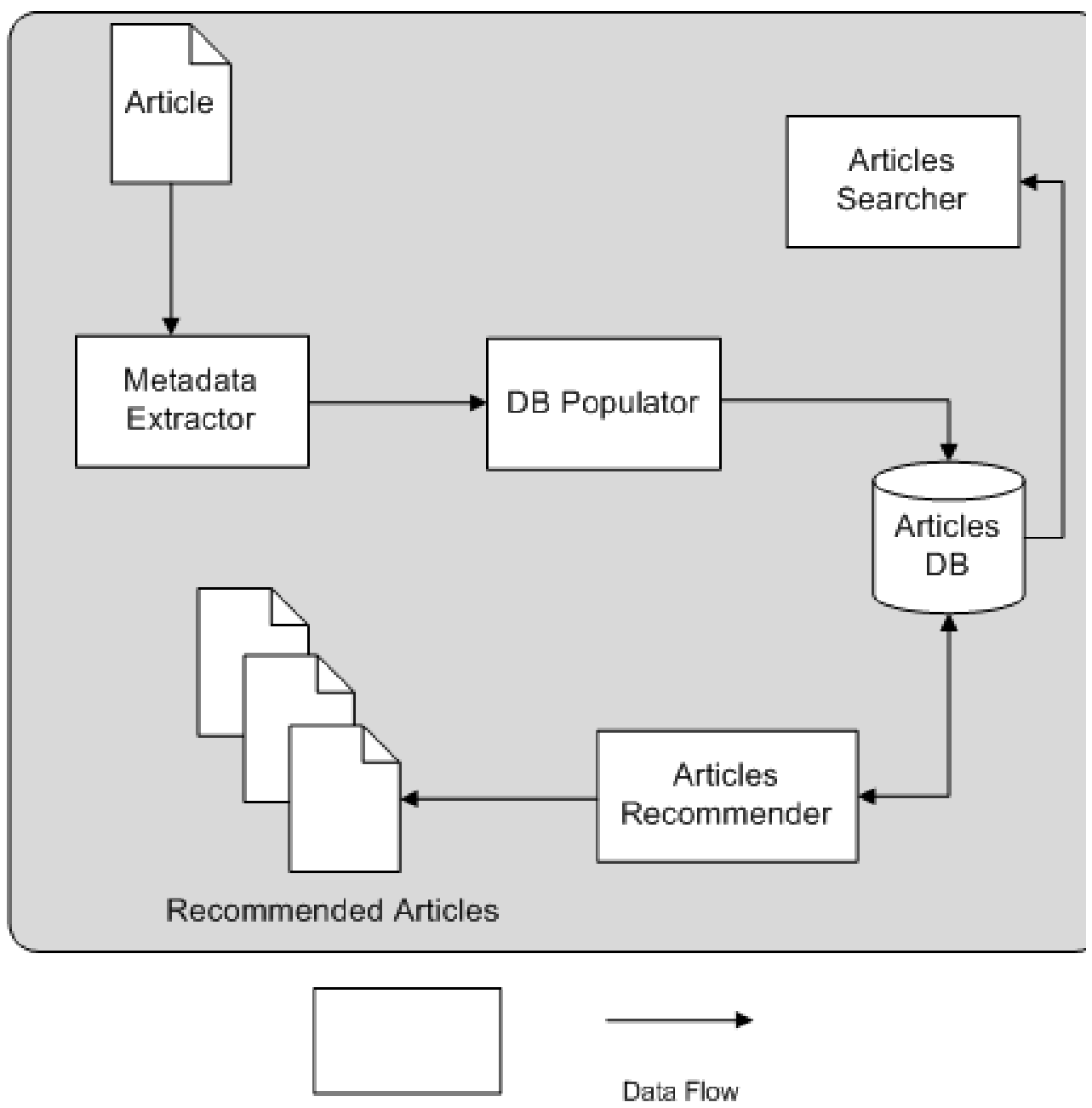

Fig. 1. Proposed system architecture

We have build our system by using Java EE 7, taking advantages of Java Persistence API (JPA) for reading and manipulating the articles metadata stored in the articles database. Of course, the initial step for using our system consists of uploading the actual articles to it. We uploaded the articles of the Bulletin of Natural Sciences in PDF format. However, there is no actual limitation from our system regarding the file type of the articles. It should be possible to support all the major file types, of course by providing an appropriate parser.

The *articles database* stores metadata about the articles, like the title, abstract, authors, keywords, body, and the article's parts. It also stores the location of the article files (PDF for the Bulletin of Natural Sciences) in the server, so that they can be provided via a user interface to the system's users. An important metadata that we store is the term frequency for each individual term, i.e. the number of times an individual term (word) appears in a single document [16]. We store the term frequency related to the actual article parts. We differentiate between "body term frequency", "title term frequency", and "abstract term frequency", and "article parts term frequency" respectively counting the frequency of a term within the body, title, abstract, and each individual part (section) of an article.

The term frequencies stored in the database are not weighted, but a term weighting scheme [16] can be derived easily by the stored frequencies, as will be shown below. We used a normalized relational database schema for storing the metadata, implemented in MySQL (using its InnoDB engine).

The *database populator* is the part of the system that stores the extracted metadata in the database. It uses the *metadata extractor* module, but it is not dependent on the actual article parser. We used JPA as an abstraction layer responsible for storing the data in the database.

The *metadata extractor* is responsible for extracting the metadata by the article files. Several parsers can be implemented according to the formatting of the articles. This module does not use any machine learning approach for detecting the article components (title, abstract, etc.) automatically from any kind of article like in [24]. We use a simpler approach, because the scope of this article is not to provide a general metadata parser from any kind of article format. In our approach for the Bulletin of Natural Sciences, we parse the PDF files of each individual article based on the formatting guidelines (text size, font weight, etc.) of them. We use *pdfbox*[3] as a PDF parser library and we extract the metadata directly from the PDF (without converting it to any other format, like text or xml). The metadata extractor makes use of an Albanian language stemmer, i.e. a software component that reduces a given word in its "stem", the part of the word that does not contain any suffixes or postfixes [16]. For example the Albanian word *bashkëpunoj* (collaborate) is reduced to *pun(ë)* (work). We used the algorithm described in [18] for stemming the words (terms). Furthermore we removed a list of stop words, words that appear most frequently in the Albanian language, based on a combined list of stop words provided by [18] and [22]. This step is critical, because the most frequent words of the Albanian language would affect in a large degree the heuristics described below that we used for finding similar articles. It is easy to change the stemming algorithm in use by the module if it is needed, this also holds for other tools used in natural language processing (NLP), like part of speech taggers described in [19] and [20].

The *articles searcher* is used when searching articles by using keyword based queries. It uses the metadata stored in the database as an index and returns search results based simply on the presence or not of a term in a particular document. The results are ranked by the frequency of the searched term (terms) in the document. The complete description of the articles searcher functionality is out of the scope of this article.

The *articles recommender* is the part of our system that behaves like a recommender system [1]. It recommends similar articles to the one that the user is currently viewing. The aim is to facilitate the discovering of articles that are about similar research questions. The similarity of articles is calculated by using a heuristics that considers the provided keywords, term frequencies located in the title, the abstract, and the body of the articles. We have implemented a content-based recommender system [5] because we lack of user profiles at this stage, and most importantly user activity information. The complete similarity calculation details are provided in the next section. This module makes use only of the articles metadata database and it is independent of the metadata extractor itself.

[3] http://pdfbox.apache.org/

For the articles recommender, metadata extractor, and articles searcher module we have also provided appropriate web services that allow for easy integration of our system with third-party articles publishing systems (i.e. other scientific journals).

The articles can be fed into the system by using the provided web service. We do not follow a web crawling approach like in [10] and [23]. The recommendation is done offline, based on the collected dataset.

## V. Similarity Calculation Heuristics

For the articles recommender module (see Figure 1) we have implemented some heuristics that calculate the similarity of two articles based on the keywords provided by the authors and the term frequencies on their respective title, abstract, and body. More concretely, in (1) we show the metrics we have used for similarity calculation. It consists of the cosine similarity of the vector space model [16], used successfully for scientific articles comparison also in [2], [10] and [23]. The similarity is calculated considering that each document is represented as a vector within a vector space whose dimensions consist of the weighted term frequencies (for each term).

$$sim(d_1, d_2) = \frac{\vec{V}(d_1)\cdot\vec{V}(d_2)}{|\vec{V}(d_1)||\vec{V}(d_2)|} = \frac{\sum_1^k w_{1i}w_{2i}}{\sqrt{\sum_1^k w_{1i}{}^2}\times\sqrt{\sum_1^k w_{2i}{}^2}} \quad (1)$$

$w_{ji}$ consists of the weighted term frequency of term $i$ in document $d_j$. To calculate the weighted term frequencies we use the heuristic presented in (2). It uses the raw term frequencies stored in the articles metadata database. As explained in section IV, the terms have been stemmed beforehand and stop words have been removed. The term frequencies of term $i$ in the keywords list, title, abstract and body of article $j$, are denoted respectively as $w^k_{ji}, w^t_{ji}, w^a_{ji}, w^b_{ji}$. We used the “terms keywords list frequency” based on the assumption that keywords chosen by the authors of an article, are the most representative terms of the content of it. $w^k_{ji}$ equals 1 if term $i$ is present in document $j$ keywords list. The title, abstract and body term frequencies have been weighted using the tf-idf scheme [15], lowering the “importance” of terms that appear too often in the whole articles’ collection.

$$w_{ji} = \kappa\, w^k_{ji} + \tau\, w^t_{ji} + \alpha\, w^a_{ji} + \beta\, w^b_{ji} \quad (2)$$

The coefficients $\kappa, \tau, \alpha$ and $\beta$ are set according to the importance of each article part (keywords, title, abstract, body) in the term frequency calculation. Because $w_{ij}$ is an affine linear combination, then $\kappa + \tau + \alpha + \beta = 1$.

Given a single article, the system calculates the similarity of it with each of the other articles of the collection. Then the results are sorted in descending similarity value order and the top $x$ similar articles are showed to the user (i.e. top 10). We generate the recommended articles by using a background job that stores the results in the articles metadata database. Due to our highly modular system architecture, it is possible to easily change the similarity function that is used to any possible similarity measure. Our testing dataset, the Bulletin of Natural Sciences, contains articles from different research categories (mathematics, computer science, biology, etc.), therefore we limited the similar document search within articles of the same category (i.e. biology).

## VI. Experiments and Evaluation

In order to test our system we ran some experiments that altered the coefficients used in (2) and the way the stemming algorithm works. The way we chose the used coefficients tried to diversify the importance of each article part based in common sense and experimentation purposes. Concretely we performed three base experiments with the following importance coefficients:

*1) $\kappa = 0.4, \tau = 0.3, \alpha = 0.2, \beta = 0.1$, giving more importance to the keywords list and title terms of an article*

*2) $\kappa = 0.0, \tau = 0.6, \alpha = 0.4, \beta = 0.0$, excluding the keywords and using only the title and abstract term frequencies*

*3) $\kappa = 0.4, \tau = 0.0, \alpha = 0.0, \beta = 0.6$, using only the keywords and body term frequencies for similarity calculation*

Furthermore we used two different stemming strategies using the algorithm proposed in [17]:

*1) We stemmed the words by using a single run (pass) of the stemming function.*

*2) We stemmed the words by using several iterations of the stemming function, until the word cannot be stemmed anymore. The ways the Albanian words are constructed creates cases that a word can be stemmed again after the first run of the stemming algorithm in use.*

It should be noted that our testing dataset contains only 226 scientific articles written in Albanian with the following distribution: 19 articles belong to the physics category, 22 to the mathematics category, 25 to the computer science category, 78 to the chemistry category, and 82 to the biology category.

Our experiments did not aim to measure the performance of the used strategy in terms of execution time, but only the relevance of the recommended articles to the input article.

We used standard evaluation measures of information retrieval systems: *precision*, *recall* and $F_1$ (a combination of precision and recall) [16], defined as in (3), (4), and (5).

$$Precision = \frac{\#(relevant\ items\ retrieved)}{\#(retrieved\ items)} = P \quad (3)$$

$$Recall = \frac{\#(relevant\ items\ retrieved)}{\#(total\ number\ of\ relevant\ items)} = R \quad (4)$$

$$F_1 = (2PR)/(P + R) \quad (5)$$

We calculated the evaluation measures for each combination of the parameters described above in this section. The experiments consisted of reviewing the recommended articles of 10 random articles of the collection. In order to find the number of relevant items, each of the recommendations was rated as "related" or "not related". This evaluation scheme is similar to the one used in [23], however we did not introduce more than one level of relevance (slightly, very, etc.). The total relevant recommendations number required in (4) was calculated by considering the total number of relevant articles found in all of the performed experiments.

TABLE I. EXPERIMENTS RESULTS

| Stemming | Coefficients used | | |
|---|---|---|---|
| | $\kappa = 0.4$ <br> $\tau = 0.3$ <br> $\alpha = 0.2$ <br> $\beta = 0.1$ | $\kappa = 0.0$ <br> $\tau = 0.6$ <br> $\alpha = 0.4$ <br> $\beta = 0.0$ | $\kappa = 0.4$ <br> $\tau = 0.0$ <br> $\alpha = 0.0$ <br> $\beta = 0.6$ |
| Single run stemming | P = 0.31 <br> R = 0.18 <br> $F_1$ = 0.23 | P = 0.34 <br> R = 0.20 <br> $F_1$ = 0.25 | P = 0.32 <br> R = 0.18 <br> $F_1$ = 0.23 |
| Multiple run stemming | P = 0.26 <br> R = 0.15 <br> $F_1$ = 0.19 | P = 0.29 <br> R = 0.17 <br> $F_1$ = 0.21 | P = 0.21 <br> R = 0.12 <br> $F_1$ = 0.15 |

The results of the experiments are displayed in Table 1. The first thing that can be noticed from the results is the fact that in general, the single run of the stemming function performs better than the multiple run. This might have happened because some words in the Albanian language may lose their real meaning when stemmed consecutively (like in our second approach).

Regarding the coefficients used in the similarity heuristics given in (2), no big differences are noticed within the experiments performed with a single run stemming. For the experiments performed with the multiple run stemming, the experiment that used only the body and keywords for calculating the weighted term frequencies, resulted the worst performing.

An interesting outcome is the fact that the weighting scheme that used only the title and abstract, performed slightly better than the ones that used also the keywords. Even though we assumed that the keywords chosen by the authors may be the most representative terms of an article, it seems that no real advantage is gained by using them for similarity calculation. This might be an indication that manually chosen keywords do not help very much a recommender system, even though they might produce good results in document classification or automated sorting.

We did not get better results by using the term frequencies of the body of the articles. This fact can be used for improving our recommender system by reducing the size of the index and also the computation time needed for calculating the term frequencies of the body, or making use of them during the similarity calculation.

Nascimento et al. [23], achieved similar results when using only the title and abstract for recommendation calculation. They also did not notice improvements when considering the term frequencies of the body of the articles.

Even though our offline calculation of the recommendations made easier the calculation of the body term frequencies in comparison with the Nascimento et al. approach, we did not gain direct benefits from it.

The evaluation scheme that we used, does not take into consideration the order of the recommended articles. Actually, when considering related items, maybe the most important factor is that the top x results contain the most relevant items. We did get fairly good results. In most of our experiments, there were many related articles in the recommendation list presented. This is an indication that the used heuristics combined with the stemming algorithm presented in [17] and the stop word lists provided in [17] and [22] can produce good results in a recommender system context. So the results showed that the stemming algorithm that was tested in a document classification context in [17], also produced good results in other information retrieval applications.

During the design of our system we also noticed some other terms that can be used as stop words, or possible improvements to the stemming algorithm. However, the tf-idf weighting of the terms reduced the importance of common words in our data set (i.e. the Albanian word sistem = system in English). It is out of the scope of this work to provide a better stemming algorithm for the Albanian language, but we believe that some custom tweaks to it that consider the most used words in a scientific domain might have produced better recommendation results.

## VII. CONCLUSIONS

Many recommender systems that help researchers on finding scientific articles related to their work have been recently proposed. There have been different approaches that usually go into the same line as recommender systems used in other areas like e-commerce, movie databases, etc. [1]. Due to the lack of the needed resources for decent information retrieval systems, there have been not much works that deal with documents written in Albanian, and even less in the scientific papers recommendation domain.

In this paper we proposed the design of a highly modular system that indexes and allows for searching of scientific articles written in Albanian. Its modular architecture simplifies the extension of it in the future, and the web services offered allow for easy integration with third-party information systems (i.e. digital libraries).

A crucial part of our designed system is the *articles recommender module*. It is a typical recommender system that recommends to a user a list of related articles (about a given single article). This facilitates a lot article searching, because after finding a particular one, other articles related to it are displayed.

In our approach we built a content-based recommender system [5] that uses cosine similarity of the vector space model for similarity calculation. Because of the similar structure of scientific articles (title, abstract, body, keywords) we used a weighting heuristics that is made of a linear combination of the term frequencies of the title, body and abstract of an article. We also used the keywords list in this heuristics, based on the assumption that keywords provided by the authors of an article, are the most descriptive terms of it.

The evaluation of our system tried different combinations of the importance factors by using a heuristic described in (2). Our goal was to identify the setting that produces better results. We also tried two different approaches of the stemming algorithm described in [17], single and multiple run of the stemming function on a given word. Our results showed that using a single run of the stemming function produced better results. Also no big differences were noticed within the results gained by the different combinations of the importance factors (coefficients) used. Nevertheless, it was showed that the body term frequencies can be excluded from the heuristics and the results will not change. This was a confirmation of the facts presented in [23].

The usage of the keywords list in the similarity heuristics did not perform better than the case that used only the title and abstract term frequencies. This fact might be an indicator that keywords defined by the authors of the article do not help very much in recommender systems scenarios.

In general, our system performed fairly well, providing relevant articles in each experiment that we made. Even though the used dataset was the one of a particular journal published in Albania, the similar structure of scientific articles allows for usage of other scientific articles, but a custom text parser need to be provided. Also, because the stemming [17] and stop word removal produced good results, we believe that similar information retrieval systems can be built in contexts not related to scientific publishing.

## VIII. FUTURE WORK

The lack of user profiles at this stage stopped us from trying a collaborative-filtering approach for our recommender system. We plan to extend the system in the future, introducing user profiles and user feedback. This way we can further improve the search experience and make another step towards an intelligent academic paper recommender system for articles written in Albanian.

We noticed possible improvements to the recommendation results if some further words have been excluded from the calculations, or have been better stemmed. We plan to test a part of speech tagger for Albanian, in order to exclude verbs in the vector space model of the articles. We believe that this might further improve the results.

The system that we built requires that articles should be uploaded to the system beforehand. A different approach would be to index articles found at the websites of other journals, and use our system only as a search engine that links to them.

Another approach that might be tried in the future is the testing of other similarity functions used in the information retrieval domain [16].

The index we created allows for further investigation of the articles dataset. It might be interesting if we can get other information from it, like research trends of the authors and topic identification [21]. Another interesting approach would be to include the authors' research trends in the recommendation formula, i.e. include articles in the recommendation list written by authors that have written articles in the topic of the article in question.

Lastly, the small dataset that we used for testing the system did not allow for careful evaluation of the performance (in terms of execution time) of the system. This will get possible when a larger dataset of articles written in Albanian will be available. At that stage we might need to tune up the system for faster recommendation generation time.

## REFERENCES

[1] G. Adomavicius, and A. Tuzhilin. "Toward the next generation of recommender systems: A survey of the state-of-the-art and possible extensions." Knowledge and Data Engineering, IEEE Transactions on 17, no. 6 (2005): 734-749.

[2] S.B. Shirude and S.R Kohle, "A library recommender system using cosine similarity measure and ontology based measure", Advances in Computational Research,Vol. 4, Issue 1, 2012, pp. 91-94.

[3] A. Tejeda-Lorente , C. Porcel, E. Peis, R. Sanz, and E. Herrera-Viedma, "A quality based recommender system to disseminate information in a University Digital Library." Information Sciences (2013).

[4] P. Lakkaraju, S. Gauch, and M. Speretta, "Document similarity based on concept tree distance.", Proceedings of the nineteenth ACM conference on Hypertext and hypermedia. ACM, 2008, pp. 127-132

[5] M.J. Pazzani and D. Billsus, "Content-based recommendation systems", in: P. Brusilovsky, A. Kobsa, W. Nejdl (Eds.), The Adaptive Web, Lecture Notes in Computer Science, vol. 4321, Springer-Verlag, 2007, pp. 325–341.

[6] J.L. Herlocker, J.A, Konstan, L.G. Terveen, and J. Riedl, "Evaluating Collaborative Filtering Recommender Systems," ACM Transactions on Information Systems, 22(1), pp. 5-53, 2004.

[7] M. Balabanovic and Y. Shoham, "Combining content-based and collaborative recommendation", Comm. ACM, vol. 40, no.3, March 1997, pp. 66 -72.

[8] J. Beel, B. Gipp, S, Langer, and M. Genzmehr, "Docear: An Academic Literature Suite for Searching, Organizing and Creating Academic Literature", Proceedings of the 11th annual international ACM/IEEE joint conference on Digital libraries (2011), 465–466.

[9] J. Beel, S. Langer, M. Genzmehr, and A. Nürnberger, "Introducing Docear's research paper recommender system", Proceedings of the 13th ACM/IEEE-CS joint conference, (JCDL '13), 2013, pp.459-460.

[10] J. Lee, K. Lee, and J. G. Kim, "Personalized Academic Research Paper Recommendation System.", arXiv preprint arXiv:1304.5457(2013).

[11] R. Torres, S. M. McNee, M. Abel, J.A. Konstan, and J. Riedl, "Enhancing Digital Libraries with Techlens", Proceedings of the 4th ACM/IEEE-CS Joint Conference on Digital Libraries (Tuscon, AZ, USA, 2004), pp. 228–236.

[12] B. Gipp, J. Beel, and C. Hentschel, "Scienstein: A Research Paper Recommender System", In Proceedings of the International Conference on Emerging Trends in Computing (ICETiC'09), Virudhunagar (India), January 2009, pp. 309–315.

[13] T. Strohman, W. B. Croft, and D. Jensen, "Recommending citations for academic papers.", In Proceedings of the 30th annual international ACM SIGIR conference on Research and development in information retrieval, SIGIR '07, pages 705–706, New York, NY, USA, 2007. ACM.

[14] T. Bogers, and A. van den Bosch, "Recommending scientific articles using citeulike", Proceedings of the 2008 ACM conference on Recommender systems (2008), 287–290.

[15] J. Tang and J. Zhang, "A discriminative approach to topic-based citation recommendation", In T. Theeramunkong, B. Kijsirikul, N. Cercone, and T.-B. Ho, editors, Advances in Knowledge Discovery and Data Mining, volume 5476 of Lecture Notes in Computer Science, pages 572–579. Springer Berlin / Heidelberg, 2009.

[16] C. D. Manning, R. Prabhaka and H. Schütze, Introduction to information retrieval, New York: Cambridge University Press, 2008

[17] J. Sadiku and M. Biba, "Automatic Stemming of Albanian Through a Rule-based Approach", Journal of International Research Publications: Language, Individuals and Society, Vol. 6, 2012

[18] Nikitas N. Karanikolas, "Bootstrapping the Albanian information retrieval.", In: BCI'09. Fourth Balkan Conference in Informatics, pp. 231-235. IEEE, 2009

[19] A. Kadriu, "NLTK tagger for Albanian using iterative approach", Information Technology Interfaces (ITI), Proceedings of the ITI 2013 35th International Conference on , vol., no., pp.283,288, 24-27 June 2013

[20] B. Hasanaj, A Part of Speech Tagging Model for Albanian An innovative solution, Saarbrücken: LAP LAMBERT Academic Publishing, 2012.

[21] D. M. Blei, "Probabilistic topic models." Communications of the ACM 55, no. 4 (2012): 77-84

[22] A. Spahiu, "100 fjalët më të shpeshta në gjuhën shqipe", May 2010, <http://www.shkenca.org/pdf/gjuhe/100_fjale.pdf>

[23] C. Nascimento, A. H. Laender, A. S. da Silva, and M. A. Gonçalves. "A source independent framework for research paper recommendation." In Proceedings of the 11th annual international ACM/IEEE joint conference on Digital libraries, pp. 297-306. ACM, 2011.

[24] R.Kern, K. Jack, M Hristakeva, and M. Granitzer, "TeamBeam Meta-Data Extraction from Scientific Literature.", D-Lib Magazine 18, no. 7, 2012

[25] C. Giles, K. Bollacker, and S, Lawrence. "CiteSeer: An automatic citation indexing system.", Proceedings of the third ACM conference on Digital libraries. ACM, 1998.